\documentclass[a4paper]{article}
\usepackage{rotating}

\usepackage[usenames]{color}
\usepackage{setspace}
\usepackage{graphicx}
\usepackage{MnSymbol}

\begin{document}

\doublespace

\title{On a mathematical relation between the Eigen model and the asexual Wright-Fisher model}
\author{Fabio Musso, Departamento de F\'isica, Universidad de Burgos\\
Facultad de Ciencias, Plaza Misael Ba\~nuelos s/n, 09001 Burgos, Spain. \\
Tel: +34947258894; Fax: +34947258831.\\ E-mail address: fmusso@ubu.es}
\date{}

\maketitle

\begin{abstract}
We show that the Eigen model and the asexual Wright-Fisher model can be obtained
as different limit cases of a unique stochastic model. This derivation makes clear which are the exact differences between these two models. 

The two key concepts introduced with the Eigen model, the error threshold and the quasispecies, are not affected by these differences, so that
they are naturally present also in population genetics models. According to this fact, 
in the last part of the paper, we use the classical diploid mutation-selection equation and the single peak fitness approximation to obtain the error threshold for sexual diploids. Finally, we compare the results with the asexual case.
\end{abstract}

\vspace{0.5cm}


\section{Introduction}

The Eigen model was formulated as a deterministic mutation-selection model describing replication at the onset of life \cite{Eigen71}.
The study of mutation-selection balance in the Eigen model for very high mutation rates led to the development of two new evolutionary concepts:
the error threshold and the quasispecies \cite{Eigen77}. The first refers to the fact that, for a critical value of the mutation probability (and for 
some choices of the fitness landscape, see \cite{Bull05}, \cite{Wilke05}, \cite{Takeuchi07}), there is an abrupt transition in the asymptotic state of the system 
from a cloud of mutants organized around 
a given consensus sequence to an almost random distribution of genotypes. The second refers to the fact that, due to the mutational coupling, 
selection acts on groups of neighbour mutants (called quasispecies) instead of individuals. 

Since RNA viruses lack proof reading mechanisms, they have mutation rates orders of magnitude higher than DNA based organisms, so that both the error threshold and the quasispecies concepts could play a relevant role for these organisms. Indeed, the Eigen model has became the main mathematical tool in this context (see \cite{Lauring10}, \cite{Mas10} for recent reviews on the subject). However, some authors questioned the relevance of the 
quasispecies as a paradigm for populations of RNA viruses \cite{Moya00}, \cite{Jenkins01}, \cite{Holmes02}, \cite{Holmes10}, suggesting that the
high heterogeneity in populations of RNA viruses could be due to genetic drift, and consequently could be better explained by population genetics models. 
This contrast could lead to the idea that the Eigen model and population genetic models are incompatible mathematical models. 
For example, \cite{Holland06} begins
saying: ``Some major differences distinguish quasispecies theory from the classical selection theories of Darwin and neo-Darwinian geneticists'', 
while in \cite{Lauring10}, one can read: ``The evolutionary dynamics of RNA viruses are complex and
their high mutation rates, rapid replication kinetics, and large population sizes present a challenge to traditional population
genetics''.
On the other hand, Wilke provided evidence that this is not the case, by showing that particular limit cases of the Eigen model give raise to 
some well known population genetics equations \cite{Wilke05}. However, the precise mathematical relation between the Eigen model and population 
genetics models remains still unclear. The main purpose of the present paper is to fill this gap, by showing that the Eigen model and 
the haploid asexual Wright-Fisher model can be obtained as different particular limits of a unique discrete time stochastic model. This is  done in section \ref{stochastic}. 
Motivated by the analogies between the Eigen model and the Wright-Fisher model, in section \ref{thresholdsex} we use the classical diploid mutation-selection equation and the single peak fitness landscape approximation, to determine the error threshold for sexual diploids. Finally, to determine the influence of syngamy on the error threshold, in section \ref{thresholdasex} we compare the results that we obtained for sexual diploids with those holding for asexual diploids.

\section{The stochastic model} \label{stochastic}

In \cite{io} it was shown that the Eigen model emerges as the deterministic and continuous time limit of a stochastic mutation-selection model. By changing the selection procedure of that model, we can obtain another stochastic model having again the Eigen model as its deterministic and continuous time limit and, at the same time, the asexual Wright-Fisher model as a different particular subcase.

Let us consider a population of constant size of $N$ individuals of $K$ possible different types $I_1,\dots,I_K$
that reproduce asexually. Let $A_i$ be the fecundity of type $I_i$, $D_i$ its degradation rate and $Q_{ij}$ the probability that an individual of type $I_j$ mutates into type $I_i$ as a result of an unexact replication
\begin{equation}
\sum_{i=1}^K Q_{ij}=1.
\end{equation}
In this model selection happens at discrete time steps $h$. Between a selection event and the successive one, the organisms of the different types $I_j$
reproduce, mutate and degradate with their characteristic rates. We assume that only the individuals in the parental generation are subject to  degradation, while the newborns will always reach the next selection step, which will  restore the total population to $N$. 

Let us denote with ${\bf{n}}=(n_1, \dots, n_K)$, $\sum_{j=1}^K n_j=N$, the type counts just after a selection event, then,
according to the above hypotheses, the expected number of individuals of type $I_i$ just before the next selection event will be given by  
\begin{equation}
m_i=n_i+h \sum_{j=1}^K \left(A_j Q_{ij} - D_i \delta_{ij} \right) n_j. \label{mi}
\end{equation} 
All the quantities appearing in the above equation, with the exception of the $n_i$ that are integer
numbers, can assume real values. Indeed, equation (\ref{mi}) should be interpreted as the 
deterministic limit of a stochastic process (see also \cite{io}).
Notice that, since in (\ref{mi}) the quantity $h D_i n_i$ represents the number of $I_i$ individuals in the parental generation that die before the next selection event, the time step length $h$ is bounded by the conditions $h D_i \leq 1$, $i=1,\dots,K$.

Selection consists in the extraction with replacement of $N$ individuals from the $\bf{m}$ population with sampling probabilities $\psi_i(\bf{n})$ equal to their relative frequencies
\begin{equation}
\psi_i({\bf{n}}):=\frac{m_i}{\sum_{j=1}^K m_j}=\frac{n_i+h\sum_{j=1}^K \left(A_j Q_{ij} - D_i \delta_{ij} \right) n_j}{N+ h \sum_{j=1}^K \left( A_j - D_j \right) n_j}. \label{psi}
\end{equation}  
Notice that, by definition, $\sum_{i} \psi_i({\bf{n}})=1$ and that $\psi_i({\bf{n}}) \geq 0$ is granted by the conditions $h D_i \leq 1$, so that the interpretation of the $\psi_i({\bf{n}})$ as probabilities is adequate. 

The Markov matrix of the model will be given by
\begin{equation}
P_h({\bf{n'}} | {\bf{n}})=\frac{N!}{n_1'! \dots n_K'!} \psi_1({\bf{n}})^{n_1'} \dots \psi_K({\bf{n}})^{n_K'}, \label{Markov}
\end{equation}
where ${\bf{n}}=(n_1, \dots, n_K)$ and ${\bf{n'}}=(n_1', \dots, n_K')$ are the type counts in successive generations, with $\sum_{j=1}^K n_j=\sum_{j=1}^K n_j'=N$. 
 
Through (\ref{psi}) and (\ref{Markov}), we have defined a family of stochastic models parametrized by the time step $h$. 
The particular case when 
\begin{equation}
h D_j=1, \ \forall j, \label{separated} 
\end{equation}
corresponds to the case of separated generations, when all the individuals in the parental generation die before the next reproductive step.
When (\ref{separated}) holds, the sampling probabilities (\ref{psi}) simplify to
\begin{equation}
\psi_i({\bf{n}}):=\frac{\sum_{j=1}^K \left(A_j Q_{ij} n_j \right)}{\sum_{j=1}^K A_j n_j}. \label{psiWF}
\end{equation}
In this case, equation (\ref{Markov}) defines the asexual Wright-Fisher model (see, 
for example, \cite{Baake00}), with the viability of type $I_i$ given by $A_i$. Notice also that the model is now 
independent of the time step $h$. We conclude that the stochastic model defined by (\ref{Markov}) reduces to the asexual haploid Wright-Fisher model when the generations are separated. 

Let us now consider the deterministic limit of the model (\ref{Markov}). First of all, let us notice that the probability that $n_i'=k$ is simply given by
\begin{equation}
P_h(n_i'=k | {\bf{n}})= 
{N \choose k}
\psi_i({\bf{n}})^{k} \left(1-\psi_i({\bf{n}})^{N-k} \right).
\end{equation}
Accordingly, the expected value of $n_i'$ will be given by
\begin{equation}
\bar{n}_i'=\sum_{k=1}^K k 
{N \choose k} 
\psi_i({\bf{n}})^{k} \left(1-\psi_i({\bf{n}})^{N-k} \right).
\end{equation}
Using the binomial identity 
\begin{equation}
\sum_{k=1}^K k  
{N \choose k} 
x^{k} y^{N-k} = N x (x+y)^{N-1},
\end{equation}
we get 
\begin{equation}
\bar{n}_i'=N \psi_i({\bf{n}}). \label{expected}
\end{equation}
By substituting the expression for the sampling probability (\ref{psi}) with $\bf{n}$ replaced by its expected value $\bar{\bf{n}}$ inside (\ref{expected}), we get the
following system of discrete equations
\begin{equation}
\bar{n}_i'=\frac{\bar{n}_i+h\sum_{j=1}^K \left(A_j Q_{ij}  - D_i \delta_{ij} \right) \bar{n}_j}{1+ h/N \sum_{j=1}^K \left( A_j - D_j \right) \bar{n}_j}. 
\label{deterministic}
\end{equation}    
Dividing equation (\ref{deterministic}) by $N$ we obtain the equations for the type frequencies $\phi_i=\bar{n_i}/N$:
\begin{equation}
\phi_i'=\frac{\phi_i+h\sum_{j=1}^K \left(A_j Q_{ij}  - D_i \delta_{ij} \right) \phi_j}{1+ h \sum_{j=1}^K \left( A_j - D_j \right) \phi_j}. 
\label{frequencies}
\end{equation}   
For $h \to 0$, we have the following asymptotic expansion
\begin{equation}
\phi_i'=\left[ \phi_i+h\sum_{j=1}^K \left(A_j Q_{ij} - D_i \delta_{ij} \right)\phi_j \right] \left\{ 1- h 
\left[ \sum_{j=1}^K \left( A_j - D_j \right) \phi_j \right] + O(h^2) \right\}, 
\end{equation}
from which it follows
\begin{equation}
\frac{\phi_i'-\phi_i}{h}= \sum_{j=1}^K \left(A_j Q_{ij}  - D_i \delta_{ij} \right) \phi_j  -\phi_i \left[ \sum_{j=1}^K \left( A_j - D_j \right) \phi_j \right] + O(h).
\end{equation}
The Eigen model equations are obtained by taking the limit $h \to 0$ of the above expression
\begin{equation}
\frac{d \phi_i}{dh}= \sum_{j=1}^K \left(A_j Q_{ij} - D_i \delta_{ij} \right)  \phi_j -
\frac{\phi_i}{N} \left[ \sum_{j=1}^K \left( A_j - D_j \right) \phi_j \right]. \label{Eigeneq}
\end{equation}
Clearly, by imposing separated generations (\ref{separated}) into equation (\ref{deterministic}) we would obtain the deterministic limit
of the haploid asexual Wright-Fisher model, that coincides with the classical haploid mutation-selection model (see, for example, \cite{Burger98}):
\begin{equation}
\phi_i'=\frac{\sum_{j=1}^K A_j Q_{ij} \phi_j}{\sum_{j=1}^K A_j \phi_j}. \label{cams}
\end{equation}
So, the only differences between the classical haploid mutation-selection model (\ref{cams}) and the Eigen model (\ref{Eigeneq}) is that the first one is obtained by the deterministic model (\ref{deterministic}) 
imposing separated generations and the latter taking the continuous time limit. 
Since neither the error threshold, nor the quasispecies phenomenon are due to these differences, 
they are naturally present in both models
(indeed, see \cite{Burger98} for the quasispecies and \cite{Wiehe95} for the error threshold in the context of population genetics).

Just to give a concrete example, let us compute the error threshold according to both models in a very special case that allows for a simple analytical
treatment. Let us suppose that each type $I_j$ is specificated by a genotype of length $L$ (so that $K=4^L$). Let $u$ be the point mutation probability and 
let us send $u \to 0$ and $L \to \infty$ in such a way that the genomic mutation rate $U=uL$ stays finite. In this limit, the probability
of mutation from the type $I_1$ to a different type $I_j, \ j \neq 1$ will be given by $\mu=1-\exp(-U)$ and the probability of back mutation
will be zero.  
Let us also consider the single peak fitness landscape $A_i=A_2, \ D_i=D_2, \ i > 2$, $A_1-D_1>A_2-D_2$. The single peak fitness landscape is a (very) simplified fitness landscape often used in the Eigen model to get an analytical expression for the error threshold (see \cite{Nowak}).
Let $\phi_B$ be the frequency of all the sequences different from $\phi_1$: 
\begin{equation}
\phi_B= \sum_{i=2}^\infty \phi_i, \label{phiB}
\end{equation} 
then the Eigen equations reduce to only two equations:
\begin{eqnarray}
\dot{\phi}_1& =& \left(A_1 e^{-U}-D_1 \right) \phi_1 - \phi_1 \left[ \left( A_1-D_1 \right)\phi_1 + \left( A_2-D_2 \right) \phi_B \right] , \label{Eigen1}\\
\dot{\phi}_B& =& A_1 \left( 1-e^{-U} \right) \phi_1  + (A_2 - D_2) \phi_M-\phi_B  \left[ \left( A_1-D_1 \right)\phi_1 + \left( A_2-D_2 \right) \phi_B \right].  \label{Eigen2}
\end{eqnarray}  
The error threshold corresponds to the smallest value of the genomic mutation rate $U$ for which
\begin{equation}
\lim_{t \to \infty} \phi_1(t)=0.
\end{equation}
From equations (\ref{Eigen1}), (\ref{Eigen2}), we get the error threshold
\begin{equation}
U_t=\ln \left( \frac{A_1}{A_2-D_2+D_1} \right).
\end{equation}
For the classical haploid mutation-selection model (\ref{cams}) the above assumptions translate in considering a locus with two alleles of relative viability $w_1=A_1>A_2=w_2$, with forward mutation given  by $\mu=1-\exp(-U)$ and the probability of back mutation being zero.
The frequencies $\phi'_1$ and $\phi'_2$ of the two alleles at the next generation, given that they are $\phi_1$ and $\phi_2$ at the present one will be:
\begin{eqnarray}
\phi'_1&=& \frac{\phi_1 A_1 (1-\mu)}{\phi_1 A_1+\phi_2 A_2},\\
\phi'_2&=& \frac{\phi_1 A_1 \mu + \phi_2 A_2}{\phi_1 A_1+ \phi_2 A_2}.
\end{eqnarray}
The first allele will go extinct, in the asymptotic limit, when $\phi'_1-\phi_1 < 0$ for $\phi_1>0$, that implies 
\begin{equation}
\mu > \frac{A_1-A_2}{A_1},
\end{equation}
or 
\begin{equation}
U_t=\ln \left( \frac{A_1}{A_2} \right).
\end{equation}
This is equivalent to the Eigen model result by keeping into account that the separated generations condition (\ref{separated}) implies that $D_1=D_2$. 
By rescaling $A_1$ to $1$ and $A_2$ to $1-s$, where $s$ is the selection coefficient, we obtain:
\begin{equation}
U_t=\ln \left( \frac{1}{1-s} \right).
\end{equation}

\section{Error threshold in the sexual diploid Wright-Fisher model} \label{thresholdsex}

Given the relation between the Eigen model (\ref{Eigeneq}) and the classical haploid mutation-selection model (\ref{cams}), it seems 
a natural option to use the classical diploid mutation-selection equation to determine the error threshold for sexual diploids.
This was indeed done in \cite{Wiehe95}, but the analytical derivation of the error threshold was inaccurate, giving the 
correct value of the critical mutation probability only for some regions of the $(h,s)$ space, where $h$ is the dominance 
and $s$ the selection parameter. To explain the problem with the derivation given in \cite{Wiehe95} let us briefly recall it.
To obtain an analytic expression for the error threshold, the authors consider a diploid analogue of the simplifying assumptions
that we used in the previous section. Namely, they considered a single locus with two alleles with mutation probability from the fittest to the worst allele given by $1-m_{11}$ 
and vanishing back mutation probability (we recall that these simplifying assumptions comes from considering a genome of infinite length in the single peak 
landscape, see the previous section). Using these assumptions, the continuous time version of the classical diploid mutation-selection equation
for the master frequency $x_1$, reduces to (compare with eq. (1) in \cite{Wiehe95}):
\begin{equation}
\dot{x}_1=x_1 (Wx)_1 m_{11}-x_1 (x,W x)\label{Wiehe1} 
\end{equation}
where (in their notations) $x=(x_1,1-x_1)$ is the vector of frequencies, $W$ the viabilities matrix and $(1-m_{11})$ is the mutation 
probability of the master sequence.
Next, they looked for a stationary solution of equation (\ref{Wiehe1}) (see eq. (17) in \cite{Wiehe95}):
\begin{equation}
x_1 (Wx)_1 m_{11}-x_1 (x,W x)=0, \label{Wiehe2}
\end{equation}
When solving equation (\ref{Wiehe2}) for $m_{11}$, they neglected the common $x_1$ factor
(see eq. (18) in \cite{Wiehe95}). That is, they solved the equation
\begin{equation}
(Wx)_1 m_{11}-(x,W x)=0 \label{Wiehe3}
\end{equation}
for $m_{11}$ in the case $x_1=0$. However, this procedure is incomplete for two reasons. First, it should be checked that 
for the obtained value of $m_{11}$ there are no more solutions of equation (\ref{Wiehe3}) for $x_1 \in (0,1]$, because in this case
$x_1=0$ would not be a global sink for the equation (\ref{Wiehe1}). 
Second, if $m_{11}$ is such that the left hand side of equation (\ref{Wiehe3}) is not zero but always negative, $x_1=0$ will be a sink
for equation (\ref{Wiehe1}) and the corresponding value of $m_{11}$ a candidate for the error threshold even if
equation (\ref{Wiehe3}) is not satisfied for $x_1=0$. So, in the following we will determine the error threshold by keeping 
into account the above considerations.   

To the effect of determining the error threshold, it is equivalent to consider the time discrete or continuous, so that we will use the  more traditional discrete time version of the classical diploid
mutation-selection equation.
As usual, we will consider two alleles on an autosomal locus in a 
monoecius random mating population with separated generations.
Let us denote by $A$ the fittest allele and by $p$ its frequency after the mutation step but before selection, while the frequency of the other allele $a$ will be given by $1-p$.
Let the relative fitness be given by $1$ for $AA$, $1-hs$ for $Aa$ and $1-s$ for $aa$. 
We will denote by $\mu$ the mutation probability
from $A$ to $a$ and, using the same approximation that we considered in the haploid case, 
we will set the back mutation probability  to zero.  
Furthermore, we will restrict our considerations to the case when $0<h<1$, that is, we will neglect underdominance and overdominance. 
The frequency $p'$ of the $A$ allele after a generation (composed by selection followed by mutation) will be given by (see \cite{Burger98}):
\begin{equation}
p'=\frac{\left(1-\mu \right) \left[ p^2+ p (1-p)(1-hs)\right]}{\bar{w}} \label{pprimo}  
\end{equation}
where $\bar{w}$ is the average fitness:
\begin{equation}
\bar{w}=p^2+2 p (1-p) (1-hs)+ (1-p)^2 (1-s)  
\end{equation}
We want to determine the minimum value of the mutation probability $\mu$ that determines the extinction of the fittest allele $A$ from the population given that its initial frequency is $p_0=1$. So, we need to find the minimum value of $\mu$ that implies $p'<p$ for any $p$. Since $\bar{w}>0$ this gives the inequality:
\begin{equation}
p (1-\mu) \left[ p + (1-p)(1-hs) \right] - \bar{w} p <0 \label{diseq}
\end{equation}
By eliminating the common $p$ factor we reduce to the quadratic inequality in $p$:
\begin{equation}
a p^2+b p +c<0 \qquad p \in (0,1], \label{diseq2}
\end{equation}
with 
\begin{eqnarray}
a&=&s(1-2h), \label{a}\\
b&=&h s (1-\mu) -2 s (1-h), \label{b}\\
c&=& (1-hs)(1-\mu)-(1-s). \label{c}
\end{eqnarray}
For $h<1/2$, both the $a$ coefficient (\ref{a}) and the discriminant $\Delta=b^2-4ac$ are positive. In this case, the parabola $a p^2+bp+c$ will always
have two real roots and will be negative in the region among the roots. One of the roots will be zero when $c=0$, that is when 
\begin{equation}
\mu=\frac{(1-h)s}{1-hs} \label{muc1}.
\end{equation}
For this value of $\mu$ the second root becomes
\begin{equation}
\frac{3 h-2+hs(1-2h)}{(1-hs)(2h-1)} \label{secondroot}.
\end{equation}
This last quantity will be greater than $1$ for $h<1$, that is always satisfied. 
We conclude that, for $h<1/2$, the error threshold is given by equation (\ref{muc1}). 
In the case $h=1/2$ the equation (\ref{muc1}) remains valid by continuity. Alternatively, since $a=0$ one can directly solve
the inequality 
\begin{equation}
bp+c<0, 
\end{equation} 
that implies 
\begin{equation}
u>\frac{1/2 s}{1-1/2 s}.
\end{equation}

When $h>1/2$, $a<0$ and the discriminant can be both positive or negative. The inequality (\ref{diseq2}) will be satisfied if one of these three
conditions is satisfied:
\begin{enumerate}
\item $\Delta \geq 0$ and the largest root is less than or equal to zero,
\item $\Delta \geq 0$ and the smallest root is greater than or equal to one,
\item $\Delta <0$.
\end{enumerate}
Equation (\ref{muc1}) implies that one of the roots is zero, so that the first condition is satisfied if the second root (\ref{secondroot})
is less than or equal to zero. This is the case if 
\begin{equation}
s> \frac{3h-2}{h(2h-1)}. \label{scondition1}
\end{equation}
Notice that the right hand side of (\ref{scondition1}) is less than zero when $1/2<h<2/3$ and it is greater than zero but less than one
when $2/3<h<1$.

The second condition can never be satisfied. Indeed, a necessary condition for the smallest root of a parabola to be greater than one is that also the 
abscissa value of the vertex $x=-b/(2a)$ be greater than one. In our case this translates into the condition
\begin{equation}
-hs(u+1)>0,
\end{equation}
that cannot be satisfied for our choice of the range of the parameters.

Regarding the third condition, there exists real $\mu$ solutions to $\Delta<0$ only when 
\begin{equation}
s<\frac{2h-1}{h^2}. \label{scondition2} 
\end{equation}
Since it holds
\begin{equation}
\frac{2h-1}{h^2} \geq \frac{3h-2}{h(2h-1)}, \qquad \frac{1}{2}<h<1,
\end{equation}
the two regions (\ref{scondition1}) and (\ref{scondition2}) cover all the region $0<s<1$, $1/2<h<1$. 
Solving $\Delta<0$ and imposing $\mu<1$, we get the solution:
\begin{equation}
\mu > \frac{2(2h-1)-h^2 s-2 \sqrt{(1-2h)(1-2h+h^2 s)}}{h^2 s}, \quad \frac{1}{2}<h<1, \quad 0<s<\frac{2h-1}{h^2}. 
\end{equation}

In the region 
\begin{equation}
{\rm{max}} \left(0,\frac{3h-2}{h(2h-1)} \right)< s < \frac{2h-1}{h^2} \qquad \frac12<h<1
\end{equation}
we have the two possible solutions for the error threshold:
\begin{eqnarray}
\mu_1&=&\frac{(1-h)s}{1-hs}\\
\mu_2&=& \frac{2(2h-1)-h^2 s-2 \sqrt{(1-2h)(1-2h+h^2 s)}}{h^2 s}
\end{eqnarray}
The two solutions $\mu_1$ and $\mu_2$ have the same value on the curve
\begin{equation}
\mu_1=\mu_2 \quad {\rm for} \quad s=\frac{3h-2}{h(2h-1)}. \label{curve}
\end{equation}
Notice that the curve for $s$ (\ref{curve}) assumes positive values only when $h>2/3$.
Since a continuous solution for the error threshold must exist in the entire region $0<h<1$, $0<s<1$, we conclude that 
the error threshold will be given by:
\begin{equation}
\left\{ \begin{array}{ll}
\mu_{\rm sex}=\mu_1 & \quad  0<h\leq \frac23, \ 0<s<1 \ {\rm or} \ \frac23<h<1, \ {\rm{max}} \left( 0,\frac{\displaystyle 3h-2}{\displaystyle h(2h-1)} \right)< s < 1\\
\mu_{\rm sex}=\mu_2 & \quad  \frac{\displaystyle 2}{\displaystyle 3}<h<1, \ s<\frac{\displaystyle 3h-2}{\displaystyle h(2h-1)}
\end{array}
\right.
\label{thresex}
\end{equation}
The result given in \cite{Wiehe95} coincides with the first line of equation (\ref{thresex}).

\section{The asexual diploids case} \label{thresholdasex}
\begin{figure}[h]
\begin{picture}(400,405)(0,0)
\put(0,0){\includegraphics[scale=0.7]{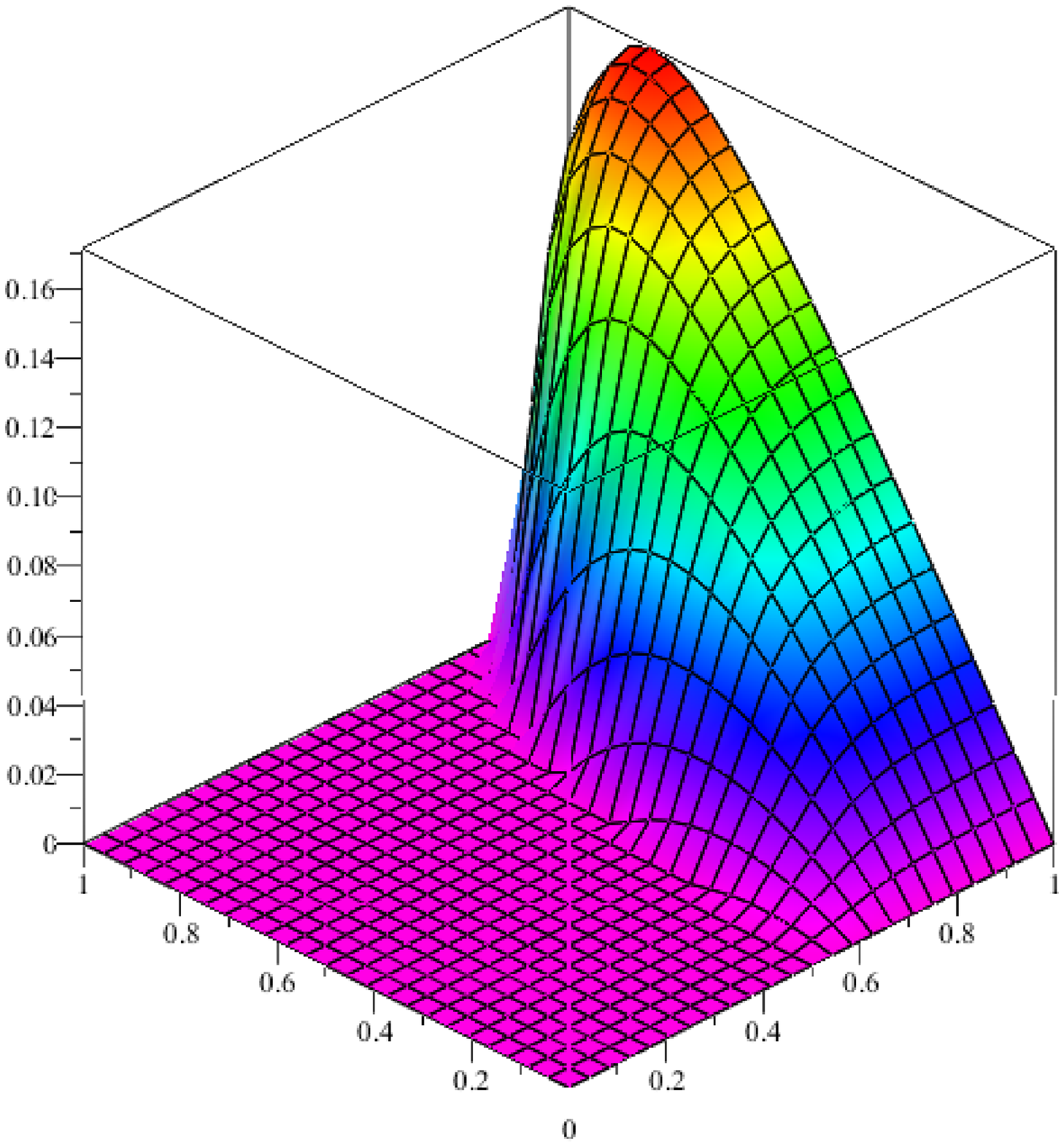} }
\put(130,65){$s$}
\put(325,75){$h$}
\put(20,220){$\mu^1_d$}
\end{picture}
\caption{The difference $\mu^1_d=\mu_{p_1=p_2=0}-\mu_{\rm sex}$ between the mutation threshold for the
complete loss of the advantageous allele in the asexual ($\mu_{p_1=p_2=0}$, see eq. (\ref{uasex})) and in the sexual case ($\mu$, see eq. (\ref{thresex})), versus the dominance $h$ and the selection coefficient $s$.}
\label{uno}
\end{figure}

It is interesting to compare the result for the error threshold of sexual diploids (\ref{thresex}) obtained in the previous section with that of asexual diploids, to evaluate the 
effect of syngamy on the error threshold. To this aim, we now calculate the error threshold for an asexual diploid organism, in the usual approximation of infinite genome length and
single peak fitness landscape.
If we have a diploid locus in an asexual organism, we denote with $p_1$, $p_2$ and $p_3$ respectively the frequencies of the $AA$, $Aa$ and $aa$ genotypes and with $\mu$ the probability of mutating from $A$ to $a$,  then under the above hypotheses, we have
that after one generation:
\begin{equation}
\vec{p}'=M \vec{p}
\end{equation}
with $M$ given by 
\begin{equation}
\left( \begin{array}{ccc}
(1-\mu)^2 & 0 &0 \\
2 \mu (1-\mu) (1-h s)& (1-\mu) (1-h s) & 0\\
\mu^2 (1-s) & \mu (1-s) &1-s
\end{array}
\right)
\end{equation}
Since the equations are linear, there is no need of normalizing. The asymptotic frequencies will be given by $p^{(\infty)}_i/(p_1^{(\infty)}+p_2^{(\infty)}+p_3^{(\infty)})$.
The outcome will depend on which is the maximum eigenvalue of the matrix $M$. If the maximum eigenvalue is $(1-\mu)^2$, then the three genotypes will coexist, because the corresponding
eigenvector of $M$ has its three components different from zero. If the maximum eigenvalue is $(1-\mu)(1-hs)$, then the $AA$ omozygote will disappear and the other two will coexist, because
the corresponding eigenvector of $M$ has the first component zero and the other two different from zero. Finally, if
the maximum eigenvalue is $1-s$, then only the homozygote $aa$ will survive, because the corresponding eigenvector of $M$ has only its third component different from zero.
The first case occurs when
\begin{equation}
\mu<{\rm min}\left(hs,1-\sqrt{1-s} \right).
\end{equation}
The second case when 
\begin{equation}
\frac{(1-h)s}{1-hs}>\mu>hs.
\end{equation}
Notice that the condition 
\begin{equation}
\frac{(1-h)s}{1-hs}>hs
\end{equation}
implies 
\begin{equation}
s>\frac{2h-1}{h^2}.
\end{equation}
Accordingly, the threshold mutation rate for the loss of the homozygote $AA$ is given by
\begin{equation}
\mu_{p_1=0}=hs \qquad {\rm max } \left( 0, \frac{2h-1}{h^2} \right) <s<1.
\end{equation}
Finally when
\begin{equation}
\mu>{\rm max}\left( \frac{(1-h)s}{1-hs}, 1-\sqrt{1-s} \right).
\end{equation}
there is the complete loss of the advantageous allele. Notice that 
\begin{equation}
 \frac{(1-h)s}{1-hs}>1-\sqrt{1-s} \quad \Leftrightarrow \quad h < \frac{1-\sqrt1-s}{s}, \label{condition} 
\end{equation}
and that 
\begin{equation}
\frac12 \leq   \frac{1-\sqrt{1-s}}{s} \leq 1, \quad 0 \leq s \leq 1. \label{range}
\end{equation}
So, the error threshold will be given by 
\begin{equation}
\mu_{p_1=p_2=0}=\left\{
\begin{array}{ll}
\frac{\displaystyle (1-h)s}{\displaystyle 1-hs} &  h < \frac{1-\sqrt{1-s}}{s},\\
1-\sqrt{1-s} &  h \geq \frac{1-\sqrt1-s}{s}.
\end{array} \right. \label{uasex}
\end{equation}
We can now compare the error thresholds in the sexual and asexual case.
Comparing equations (\ref{thresex}), (\ref{uasex}) and keeping into account (\ref{condition}) and (\ref{range}), we see that for $h \leq 1/2$
there is no difference in the error threshold between the sexual and asexual case. We plot this difference in figure \ref{uno} for the whole
range of variation of $h$ and $s$.
We see that, for $h > 1/2$ , the advantageous allele is more robust to complete loss by mutation in the asexual than in the sexual case.

\begin{figure}[h]
\begin{picture}(400,405)(0,0)
\put(0,0){\includegraphics[scale=0.7]{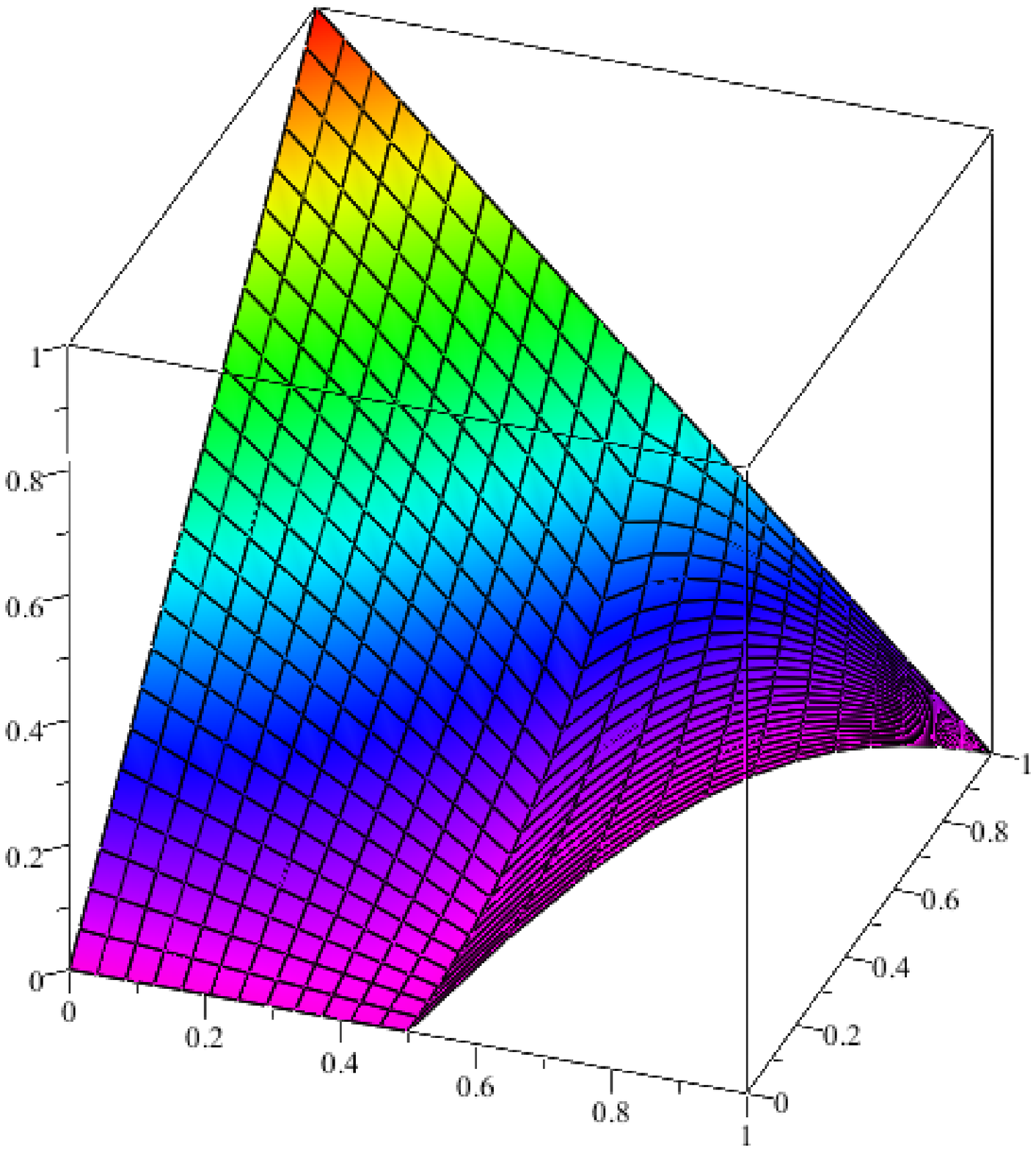} }
\put(175,50){$h$}
\put(355,100){$s$}
\put(40,200){$\mu^2_d$}
\end{picture}
\caption{The difference $\mu^2_d=\mu_{\rm sex}-\mu_{p_1=0}$ between the mutation threshold for the
loss of the advantageous homozygote in the sexual ($\mu$, see eq. (\ref{thresex})) and in the asexual case ($\mu_{p_1=0}$, see eq. (\ref{uasex})), versus the dominance $h$ and the selection coefficient $s$.}
\label{due}
\end{figure}

In figure \ref{due} we show the difference between the sexual error threshold (\ref{thresex}) and the threshold mutation rate for the loss
of the advantageous homozygote in the asexual case. 
We see that in this case, as obvious, the advantageous homozygote is much more robust to loss by mutation in the sexual than in the asexual case. Indeed, in the sexual case the advantageous homozygote can be eliminated only by completely removing the advantageous 
allele.

\section{Conclusions}

We constructed a stochastic model having the haploid Wright-Fisher model and the Eigen model as particular subcases. The haploid Wright-Fisher model is obtained by considering separated generations, while the Eigen model is obtained by taking the deterministic and continuous time limit. This derivation makes it clear what are the differences between these two important models 
of mutation-selection dynamics. Emerging as a deterministic limit, the Eigen model neglects genetic drift and it is almost equivalent to the deterministic limit of the haploid Wright-Fisher model,
that is, the classical haploid mutation-selection model (\ref{cams}). The differences among this model and the Eigen model do not invalidate the concepts of quasispecies and error threshold, that
consequently are present in both models.  This suggests to use the classical diploid mutation-selection model to obtain the error threshold for sexual diploids. We derived an analytical expression for the error threshold inside this model by using the usual approximations of infinite 
genome length and the single peak fitness landscape. We compared this expression with
the corresponding expression for asexual diploid organisms. No difference emerges when $h \leq 1/2$, but, curiously, when $h>1/2$, syngamy makes the advantageous allele more liable to complete loss by mutation. 
On the other hand, this is not the case for the loss of the advantageous 
homozygote that in the sexual case, especially for low values of the dominance parameter $h$ and high values of the selection coefficient as can be appreciated in figure

\singlespace

\newpage

\newpage

\end{document}